\documentclass{amsart}

\newtheorem{theorem}{Theorem}[section]

\newtheorem{proposition}[theorem]{Proposition}

\theoremstyle{definition}

\theoremstyle{remark}
\newtheorem{remark}[theorem]{Remark}

\numberwithin{equation}{section}

\begin{document}

\title[A New Efficient Asymmetric Cryptosystem Based on the Integer Factorization Problem] {A New Efficient Asymmetric Cryptosystem Based on the Integer Factorization Problem}

%Information for first author
\author{M.R.K. Ariffin}
\address{Al-Kindi Cryptography Research Laboratory, Institute for Mathematical Research,
Universiti Putra Malaysia, 43400 UPM,Serdang, Selangor, MALAYSIA}
\curraddr{Department of Mathematics, Faculty of Science, Universiti
Putra Malaysia, 43400 UPM, Serdang, Selangor, MALAYSIA}
\email{rezal@putra.upm.edu.my}
\thanks{The research was supported by the Fundamental Research Grant Scheme
$\#5523934$ and Prototype Research Grant Scheme $\#5528100$ Ministry
of Higher Education, MALAYSIA.}

%\author{M.A.Asbullah}
%\address{Al-Kindi Cryptography Research Laboratory, Institute for Mathematical Research,
%Universiti Putra Malaysia, 43400 UPM,Serdang, Selangor, MALAYSIA}

%\curraddr{Centre of Foundation Studies for Agricultural Science,
%Universiti Putra Malaysia, 43400 UPM, Serdang, Selangor, MALAYSIA}
%\email{ma$\_$asyraf@putra.upm.edu.my}
%\thanks{Ministry of Higher Education, MALAYSIA}

%\author{N.A. Abu}
%\address{Al-Kindi Cryptography Research Laboratory, Institute for Mathematical Research,
%Universiti Putra Malaysia, 43400 UPM,Serdang, Selangor, MALAYSIA}
%\curraddr{Department of Computer Systems and Communication,
%Faculty of Information and Communication Technology, Universiti
%Teknikal Malaysia Melaka, 76109 Durian Tunggal, Melaka, MALAYSIA}
%\email{nura@utem.edu.my}

%\author{Z. Mahad}
%\address{Al-Kindi Cryptography Research Laboratory, Laboratory for Theoretical Studies,
%Institute for Mathematical Research, Universiti Putra Malaysia, 43400 UPM,
%Serdang, Selangor, MALAYSIA}
%\email{putra_ikim@gmail.com}

%\author{M.A. Daud}
%\address{Al-Kindi Cryptography Research Laboratory, Laboratory for Theoretical Studies,
%Institute for Mathematical Research, Universiti Putra Malaysia, 43400 UPM,
%Serdang, Selangor, MALAYSIA}
%\email{putra_ikim@gmail.com}

%    General info

%\date{January 1, 2001 and, in revised form, June 22, 2001.}



\begin{abstract}
A new asymmetric cryptosystem based on the Integer Factorization
Problem is proposed. It posses an encryption and decryption speed of
$O(n^2)$, thus making it the fastest asymmetric encryption scheme
available. It has a simple mathematical structure. Thus, it would
have low computational requirements and would enable communication
devices with low computing power to deploy secure communication
procedures efficiently.
\end{abstract}

\maketitle

%\section*{This is an unnumbered first-level section head}
%This is an example of an unnumbered first-level heading.

%% The correct journal style for \specialsection is all uppercase; a known bug
%% in amsart.cls prevents this, so input must be uppercase until it is fixed.
%\specialsection*{This is a Special Section Head}
%\specialsection*{THIS IS A SPECIAL SECTION HEAD}
%This is an example of a special section head%
%%%%%%%%%%%%%%%%%%%%%%%%%%%%%%%%%%%%%%%%%%%%%%%%%%%%%%%%%%%%%%%%%%%%%%%%
%\footnote{Here is an example of a footnote. Notice that this footnote
%text is running on so that it can stand as an example of how a footnote
%with separate paragraphs should be written.
%\par
%And here is the beginning of the second paragraph.}%
%%%%%%%%%%%%%%%%%%%%%%%%%%%%%%%%%%%%%%%%%%%%%%%%%%%%%%%%%%%%%%%%%%%%%%%%

\section{Introduction}
By textbook convention the discrete log problem (DLP) and the
elliptic curve discrete log problem (ECDLP) has been the source of
security for cryptographic schemes such as the Diffie Hellman key
exchange (DHKE) procedure, El-Gamal cryptosystem and elliptic
curve cryptosystem (ECC) respectively \cite{diffie},
\cite{koblitz}. As for the world renowned RSA cryptosystem, the
inability to find the $e$-th root of the ciphertext C modulo N
from the congruence relation $C\equiv M^e (\textrm{mod }N)$
coupled with the inability to factor $N=pq$ for large primes $p$
and $q$ is its fundamental source of security \cite{rsa}. It has
been suggested that the ECC is able to produce the same level of
security as the RSA with shorter key length. Thus, ECC should be
the preferred asymmetric cryptosystem when compared to RSA
\cite{vanstone}. Hence, the notion ``cryptographic efficiency" is
conjured. That is, to produce an asymmetric cryptographic scheme
that could produce security equivalent to a certain key length of
the traditional RSA but utilizing shorter keys. However, in
certain situations where a large block needs to be encrypted, RSA
is the better option than ECC because ECC would need more
computational effort to undergo such a task \cite{scott}. Thus,
adding another characteristic toward the notion of ``cryptographic
efficiency" which is it must be less ``computational intensive"
and be able to transmit large blocks of data (when needed). In
1998 the cryptographic scheme known as NTRU was proposed with
better "cryptographic efficiency" relative to RSA and ECC
\cite{hoffstein2} \cite{hermans} \cite{hoffstein3}. NTRU has a
complexity order of $O(n^{2})$ for both encryption and decryption
as compared to DHKE, EL-Gammal, RSA and ECC (all have a complexity
order of $O(n^{3})$). As such, in order to design a
state-of-the-art public key mechanism, the following are
characteristics that must be ``ideally" achieved (apart from other
well known security issues):

\begin{enumerate}
    \item Shorter key length. If possible shorter than ECC 160-bits.
    \item Speed. To have speed of complexity order $O(n^2)$ for both encryption and decryption.
    \item Able to increase data set to be transmitted asymmetrically. That is, not to be restricted in size because of the mathematical structure.
    \item Simple mathematical structure for easy implementation.
\end{enumerate}

In this paper, we produce a newly designed asymmetric cryptosystem
based on the Integer Factorization Problem. The scheme does not
require ``expensive" operations. It only requires multiplication and
addition for encryption and for decryption it only utilizes
multiplication together with a one time modular
reduction.\\

The layout of this paper is as follows.  In Section 2, we define the
Diophantine Equation Hard Problem (DEHP) which is the source of
``mathematical hardness" integrated within the ciphertext equation.
The new asymmetric cryptosystem will be detailed in Section 3. In
Section 4, the authors detail the decryption process and provide a
proof of correctness. An example will also be presented. Continuing
in Section 5, we will discuss algebraic attacks. An analysis of
lattice based attack will be given in Section 6. Section 7 will be
about the underlying security principles of the $AA_{\beta}$ scheme.
A table of comparison between the $AA_{\beta}$ scheme against
RSA,ECC and NTRU is given in Section 8. Finally, we shall conclude
in Section 9.

\section{The Diophantine Equation Hard Problem (DEHP)}

In this section we begin by producing a diophantine equation of the
form
$$
C=Ax+By
$$
where $(A,B,C)$ are known integers while $(x,y)$ are unknown
integers. Another condition is that gcd$(A,B)=1$. We will observe
the following 2 cases:

\subsection{Case 1}
Let the public parameters $(A,B)$ be of length $n$-bits and the
secret parameters $(x,y)$ also be of length $n$-bits.
%The equation
%given by $C=Ax+By$ is of length $2n$-bits. In obtaining the
%initial points $(x_{0},y_{0})$ via the extended euclidean
%algorithm, we can observe that the size of $(x_{0},y_{0})$ is
%equal or greater $2n$-bits.

The general solution for $(x,y)$ is given by
\begin{itemize}
    \item $x=x_{0}+Bt$
    \item $y=y_{0}-At$
\end{itemize}
Since the size if the unknown parameters $(x,y)$ are $n$-bits, from
the following inequality
$$
\frac{2^{n-1}-x_{0}}{B}<t<\frac{2^{n}-1-x_{0}}{B}
$$
and by the fact that $2^{n-1}<B<2^{n}-1$, the interval that the
variable $t$ belongs to is approximately given by
$$
\frac{2^{n}}{B}>\frac{2^{n}}{2^{n}-1}\approx 1
$$
As a result an attacker could be able to determine the value of $t$
and solve for the unknown pair $(x,y)$ in polynomial time.

\subsection{Case 2}
Let the public parameters $(A,B)$ be of length $n$-bits and the
secret parameters $(x,y)$ be of length $2n$-bits. The general
solution for $(x,y)$ is given by
\begin{itemize}
    \item $x=x_{0}+Bt$
    \item $y=y_{0}-At$
\end{itemize}
Since the size if the unknown parameters $(x,y)$ are $2n$-bits, from
the following inequality
$$
\frac{2^{2n-1}-x_{0}}{B}<t<\frac{2^{2n}-1-x_{0}}{B}
$$
and by the fact that $2^{n-1}<B<2^{n}-1$, the interval that the
variable $t$ belongs to is approximately given by
$$
\frac{2^{2n}}{B}>\frac{2^{2n}}{2^{n}-1}\approx 2^{n}
$$
As a result an attacker could not be able to determine the value of
$t$ and solve for the unknown pair $(x,y)$ in polynomial time for
sufficiently large $n$.

\subsection{Definition (DEHP)} The Diophantine Equation Hard Problem (DEHP)
is the problem to determine the preferred solution set $(x,y)$ from
$C=Ax+By$ where $(A,B,C)$ in known and $(x,y)$ is unknown. A correct
implementation of DEHP will be executed as describe in Case 2 above.
If the preferred solution set is obtained then the equation $C$ is
said to be $prf$-solved.

\section{A new asymmetric algorithm based on Integer Factorization Problem}

Let us begin by stating that the communication process is between
A (Along) and B (Busu), where Busu is sending information to Along
after encrypting the plaintext with Along's public key.\\

$\bullet$ \textbf{Key Generation by Along}\\
\newline
\indent INPUT: Generate a pair of random $n$-bit prime numbers $p$
and $q$, an $n$-bit odd integer $k_1$, $k_2=\frac{q-k_{1}}{2}$ and a
random $2n$-bit integer $u$. Another condition is that $p>2^{n-1}+2^{n-2}$.\\
\indent OUTPUT: The public key $e_{1}$ and $e_{2}$ where
\begin{itemize}
    \item $e_{1}=u+p(k_{1}+k_{2})$
    \item $e_{2}=u-pk_{2}$
\end{itemize}
and the private key pair $(p,d)$ where $d\equiv v^{-1}(\textrm{mod }p)$ and $v\equiv u (\textrm{mod }p)$.\\

$\bullet$ \textbf{Encryption by Busu}\\
\newline
\indent INPUT: The public key $(e_{1},e_{2})$ and the message $M$
where $M$ is an $n$-bit integer within the interval
$(2^{n-1},2^{n}-1)$ and $M<2^{n-1}+2^{n-2}$. As a result $M<p$.
Compute a random $3n$-bit integer $X$ and compute $Y=X-M$.\\
\indent OUTPUT: The ciphertext $C=Xe_{1}-Ye_{2}$.\\
%\newpage

$\bullet$ \textbf{Decryption by Along}\\
\newline
\indent INPUT: The private key pair $(p,d)$ and the ciphertext $C$. \\
\indent OUTPUT: The plaintext $M$.

\section{Decryption}

\begin{proposition}
$Cd\equiv M(\textrm{mod }p)$.
\end{proposition}

\noindent We now proceed to give a proof of correctness.
\newline

\begin{proof}
$Cd\equiv X-Y \equiv M(\textrm{mod }p)$. Observe that, modular
reduction does not occur since $M<p$.
\end{proof}

\subsection{Example} Let $n=16$. Along will choose the primes
$p=65287$ and $q=40829$. Then Along chooses the following private
parameters:
\begin{enumerate}
\item $k_{1}=46381$
\item $k_{2}=-2776$
\item $u=3096817651$
\item $d=49913$
\end{enumerate}

The public keys will be
\begin{enumerate}
\item $e_{1}=5943657286$
\item $e_{2}=3278054363$
\end{enumerate}

Busu's message will be $M=43963$ with the following accompanying
parameters
\begin{enumerate}
\item $X=281474976710656$
\item $Y=281474976666693$
\end{enumerate}
The ciphertext will be $C=750300520815394662808057$. To decrypt is
arbitrary. $_\Box$

 \section{Algebraic Attacks}
\vspace{0.25cm}

\subsection{Computing with X}

\noindent To find $X=X_{0}+e_{2}j$, we should find an integer $j$
such that $2^{3n-1}<X<2^{3n}-1$. This gives
$$\frac{2^{3n-1}-X_{0}}{e_{2}}<j<\frac{2^{3n}-1-X_{0}}{e_{2}}.$$ We know that $2^{2n-1}<e_{2}<2^{2n}-1.$
Then the difference between the upper and the lower bound is
$$\frac{2^{3n}-1-X_{0}}{e_{2}}-\frac{2^{3n-1}-X_{0}}{e_{2}}\approx\frac{2^{3n}}{e_{2}}>\frac{2^{3n}}{2^{2n}-1}\approx2^{n}.$$
Hence the difference is very large and finding the correct $j$ is infeasible.\\

\begin{remark}
If one attempts to compute with $Y$ the above scenario when
computing with $X$ would appear.
\end{remark}

\subsection{Euclidean division attack} From $C=Xe_{1}-Ye_{2}$,
the size of each public parameter within $C$ ensures that
Euclidean division attacks does not occur. This can be easily
deduced as
follows:\\
\begin{enumerate}
\item $\lfloor\frac{C}{e_{1}}\rfloor \neq X$
\item $\lfloor\frac{C}{e_{2}}\rfloor \neq Y$\\
\end{enumerate}

\section{Analysis on lattice based attack}

With reference to the $AA_{\beta}$ scheme in \cite{ariffin} which
has gone through square lattice attack, the ciphertext equation in
this article is of the same structure. Recall that the the structure
of the ciphertext $AA_{\beta}$ scheme is as follows:
$$
C=Ue_{A1}+V^{2}e_{A2}
$$
where both $(U,V^{2})$ are of size $4n$-bits, while both
$(e_{A1},e_{A2})$ are of size $3n$-bits. That is, the unknown
parameters are larger by $n$-bits from the known parameters. It is
by this fact that the square lattice attack as described in
\cite{ariffin} failed upon the $AA_{\beta}$ scheme.\\

Now, observe that the ciphertext in this article also has its
unknown parameters to be larger by $n$-bits than the known
parameters. It is arbitrary to replicate the empirical evidence by
executing the LLL algorithm upon the scheme in this article to see
that the square lattice attack will not succeed.

\section{Underlying security principles}

\subsection{The public key Integer Factorization Problem}
%$\bullet$ \underline{The integer factorization problem}\\

Observe that one can obtain $e_{1}-e_{2}=pq$. This is obviously the
Integer Factorization Problem.

\subsection{The ciphertext DEHP}

To find the preferred solution set $(X,Y)$ such that
$C=Xe_{1}-Ye_{2}$.

\section{Table of Comparison}

The following is a table of comparison between RSA, ECC, NTRU and
the scheme in this article. Let $|E|$ denote public key size.\\

\begin{center}
\begin{tabular}{|c|c|c|c|c|}
  \hline
  Algorithm & Encryption & Decryption & Ratio  & Ratio \\
   & Speed & Speed & $M:C$ & $M:|E|$  \\
  \hline
  RSA & $O(n^3)$ & $O(n^3)$ & $1:1$ & $1:2$ \\
  \hline
  ECC & $O(n^3)$ & $O(n^3)$ & $1:2$ & $1:2$ \\
  \hline
  NTRU & $O(n^2)$ & $O(n^2)$ & Varies \cite{hoffstein2} & N/A \\
  \hline
  Scheme in this paper & $O(n^2)$ & $O(n^2)$ & $1:5$ & $1:4$  \\
  \hline
\end{tabular}
\end{center}

\begin{center}
{Table 1. Comparison table for input block of length $n$}
\end{center}

\section{Conclusion}

The asymmetric scheme presented in this paper provides a secure
avenue for implementors who need encryption and decryption speed of
complexity order $O(n^2)$.

% do the biliography:

\bibliographystyle{amsplain}

\end{document}